\documentclass[journal]{IEEEtran}
\usepackage{cite}
\usepackage{amsmath,amssymb,amsfonts}
\usepackage{algorithmic}
\usepackage{graphicx}
\usepackage{textcomp}
\usepackage{xcolor}
\usepackage{subfig}
\usepackage{url}

\newcommand{\eg}{\textit{e.g.~}}
\newcommand{\ie}{\textit{i.e.~}}

\DeclareMathOperator*{\argmax}{argmax}

\newcommand{\etc}{\textit{etc.}}

\newcommand\ignore[1]{}

\newcommand{\figref}[1]{Fig.~\ref{#1}}

\newcommand{\cf}{\textit{cf.~}}

\def \path{\bp C}

\newcommand{\bfH}{{\mathbf{H}}}

\newcommand{\bfR}{{\mathbf{R}}}

\newcommand{\calG}{{\mathcal{G}}}
\newcommand{\calH}{{\mathcal{H}}}
\newcommand{\calI}{{\mathcal{I}}}

\newcommand{\calP}{{\mathcal{P}}}

\newcommand{\bbN}{{\mathbb{N}}}

\newcommand{\bbR}{{\mathbb{R}}}

\begin{document}

\title{Static Visual Spatial Priors for DoA Estimation}

\author{Pawel Swietojanski$^*$,~\IEEEmembership{Member,~IEEE,}
        and~Ondrej~Miksik$^*$
\thanks{$^*$ Equal contribution.}
\thanks{P. Swietojanski is with the School of Computer Science and Engineering, The University of New South Wales, Australia. (p.swietojanski@unsw.edu.au)}
\thanks{O. Miksik is with Emotech Labs, UK.}
}

%
%


\maketitle

\begin{abstract}
As we interact with the world, for example when we communicate with our colleagues in a large open space or meeting room, we continuously analyse the surrounding environment and, in particular, localise and recognise acoustic events.
While we largely take such abilities for granted, they represent a challenging problem for current robots or smart voice assistants as they can be easily fooled by high degree of sound interference in acoustically complex environments.
Preventing such failures when using solely audio data is challenging, if not impossible since the algorithms need to take into account wider context and often \emph{understand} the scene on a \emph{semantic level}.
In this paper, we propose what to our knowledge is the first multi-modal \emph{direction of arrival} (DoA) of sound, which uses \emph{static visual spatial prior} providing an auxiliary information about the environment to suppress some of the false DoA detections.
We validate our approach on a newly collected real-world dataset, and show that our approach consistently improves over classic DoA baselines. 
\end{abstract}


\vspace{-0.3cm}
\section{Introduction}

\IEEEPARstart{D}{irection of Sound Arrival} (DoA)~\cite{dibiase2000, Schmidt1986, Yoon2006}, or in general acoustic source localisation, played an important role in recent widespread adoption of voice assistants, in particular for devices that are more designed like robots with some degree of freedom in the environment. 
DoA is typically used for improving spatial scene understanding and as such forms basis for decision making, \eg to take specific physical actions (rotate to the user, steer to an object of interest, \etc).
Thus, it plays an important role in the overall user experience. 

However, sound source localisation often becomes inherently ambiguous whenever the acoustic environment gets more complex.
Consider, for instance, a single sound source and a device (microphone array) placed next to a glass wall; strong sound reflections from the wall often lead to unwanted interference that confuses DoA estimates. 
Rotating the robot to such location instead to the user can entirely break the user experience. 
But how can we \emph{identify} the \emph{true} sound source? 
Often, it is difficult or even impossible to disambiguate between the two using raw audio signal alone without any understanding of the wider context or having auxiliary prior knowledge about expected behaviour (\eg estimate DoAs of people in the room, rather than DoAs of noises on the street).

\begin{figure}[t]
    \centering
    \includegraphics[width=0.475\linewidth]{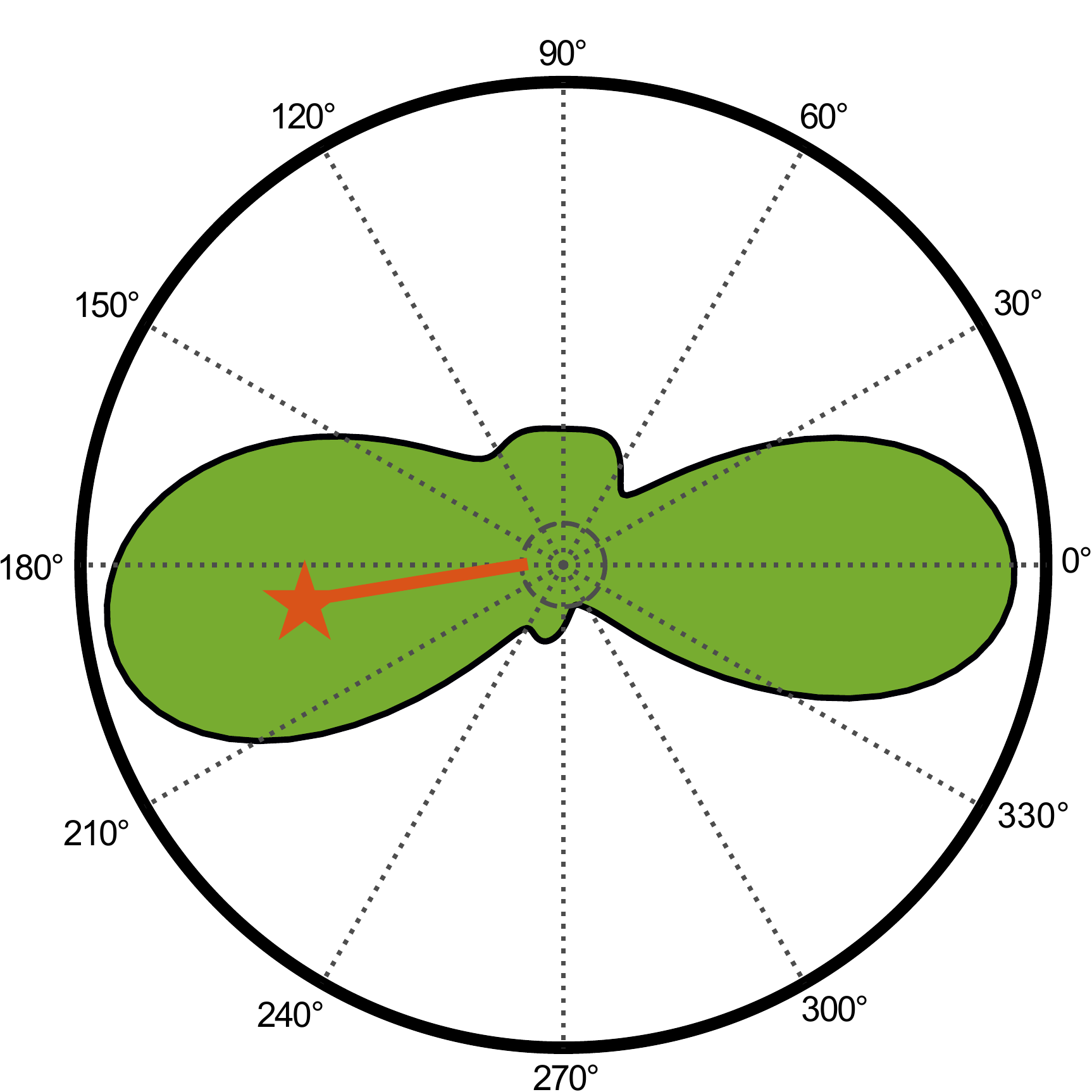}
    \hfill{}
    \includegraphics[width=0.475\linewidth]{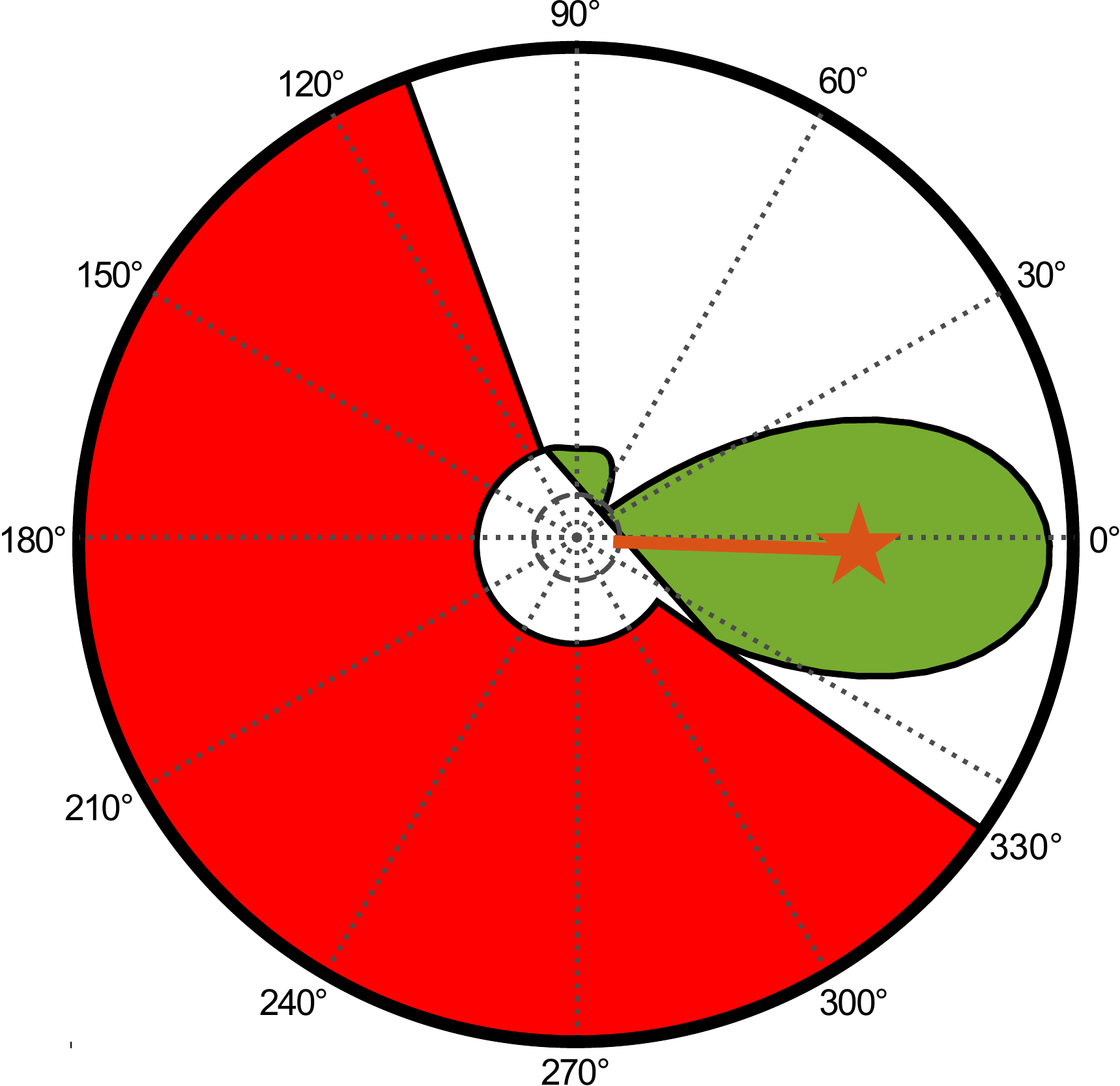}
    \vspace{-0.4cm}
    \definecolor{spectrum}{RGB}{119,172,48} 
\definecolor{reconstruction}{RGB}{217,83,25} 
\definecolor{prior}{RGB}{255,0,0}

{
\begin{tabular}{lll}
\colorbox{spectrum}{\textcolor{spectrum}{|}} spectrum~~&
\colorbox{reconstruction}{\textcolor{reconstruction}{|}} estimated DoA~~&
\colorbox{prior}{\textcolor{prior}{|}} visual prior
\end{tabular}
} 
    \vspace{-0.25cm}
    \caption{(left) DoA is often confused in acoustically complex environments. (right) While identifying the true DoA using acoustic data alone may be challenging, injecting the algorithm with \emph{static visual prior} provides an auxiliary information about the environment, that can be used to suppress false detections.
    Ground-truth prediction for this example is $0^\circ$.
    }
    \label{fig:teaser}
    \vspace{-0.3cm}
\end{figure}

Multi-sensory and multi-modal processing (sensors capture information from different origins) has been found to greatly improve performance in various machine learning perception tasks, in particular the ones with inherent ambiguity.  
Thus, it is viable to provide the DoA models with a spatial prior describing the vicinity of the device.
With recent advances in computer vision, visual modality makes a good candidate for building such a prior since i) it can provide detailed information about the environment~\cite{valentin2013cvpr, vineet2015icra, davison2018spatialai} and ii) majority of potential multimedia hardware are equipped with cameras, hence such algorithms come with no extra (hardware) cost.

In this paper, we propose what to our knowledge is the first multi-modal DoA which uses \emph{static visual spatial priors} to suppress false DoA detections (\cf \figref{fig:teaser}).
Specifically, we assume a hardware platform equipped with a microphone array and a standard monocular camera mounted on a moving head which can rotate around its vertical axis.
Using this platform, we capture $360^{\circ}$ images which we use to predict \emph{free space} and \emph{obstacles}, \ie to estimate plausible regions that can be occupied by people.
To this end, we use semantic image segmentation as a proxy for sound source regions of interest.
Then, we inject such static visual prior into classic audio-based DoA algorithms and show that it significantly reduces errors. 
We focus on injecting \emph{static priors}, which is orthogonal to most multi-modal approaches that typically consider simultaneous and (often) synchronized data streams.  
The key difference is, that the static prior is estimated infrequently (\ie calibration stage), when compared to the information throughput of the \emph{primary} audio data stream (always on). This allows to use visual information with a constant compute cost, and in a broader set of situations -- our approach is not sensitive to performance degradation in low-light conditions, nor assumes the users to be present within the the camera field-of-view. Additionally, it does not require specialised hardware with active depth sensing (though in general could use it).

\section{Related work}

\vspace{-0.02cm}

\textbf{Direction of arrival}. Estimating direction of arrival (DoA) of sound requires an access to a multi channel source of acoustic signal, typically captured by an $M$-element microphone array of (an ideally) known geometry. DoA can be then estimated directly by time-aligning the signals captured by pairs of microphones, using for example, Generic Cross Correlation with Phase Transform (GCC-PHAT) approach~\cite{Knapp1976, Brandstein1997}. Computing pair-wise DoAs in time delay domain, however, does not allow to fully utilise redundant information that results from combining several microphones in signal domain (and which is important in more challenging acoustic environments). Steered Response Power with PHAT (SRP-PHAT)~\cite{dibiase2000} scans through candidate directions and seeks for the peaks in the value of GCC-PHAT averaged over all microphone pairs, this value is assumed as a desired DoA. 

Signal sub-spaces methods estimate the so called \emph{spatial covariance matrix} between multiple channels~\cite{Schmidt1986} and assume that the true and noise sources are independent and uncorrelated, thus occupy different subspaces. 
%
%
This has been generalised using maximum likelihood~\cite{Stoica1990, Pesavento2001}, by exploiting statistical regularities \ie \emph{test of orthogonality of projected subspaces} (TOPS) \cite{Yoon2006} or weighting subspaces when deriving DoAs~\cite{Claudio2001}.

Robust estimation of DoA in reverberant environments remain an active research area, some directions include techniques for smoothing DoA trajectories~\cite{Anguera2007, rafaely2017speaker, Evers2017} or the use of distributed sensors \cite{Canclini2013, Evers2018}. The former approach is hard to apply in low-latency settings, while the latter is not always practical, though our method remains complementary to either. Recently, a trainable  neural-net-based DoA was proposed \cite{adavanne2018direction}.

Using audio-visual information to improve speaker localisation and tracking has also been studied, \eg \cite{strobel2001joint} or \cite{gehrig2005kalman} used parallel audio-visual information for speaker tracking. It also remains an active research area in acoustic SLAM~\cite{Debru2017}. However, those approaches assume parallel data-streams and processing, whereas our work is concerned with independent asynchronous and non-real-time injection of visual data. 

%
\textbf{Visual \emph{free space} prediction}. Detecting \emph{free space} from visual data has been widely explored in robotics.
Free space is usually characterised by semantic classes corresponding rather to \emph{stuff} than \emph{objects}~\cite{adelson2001thigsstuff}, hence this task is typically formulated as dense labelling (instead of using sparse representations such as bounding boxes) where the goal is to assign a (binary) label corresponding to free space or obstacles to each pixel.
In robotics, semi-supervised methods used LIDARs or stereo-cameras to propose weak labels~\cite{dahlkamp2006, cnn4road}.
Later, this was adapted to a single monocular camera~\cite{miksik2011icra}.
Modern approaches go beyond binary labelling and rather segment the scene into semantically meaningful regions (\eg floor, wall, TV, \ldots), typically using multi-class structured prediction frameworks~\cite{ahcrf, dsm, crfasrnn2015iccv}.
The main advantages of multi-class segmentation is that i) it provides more information about the environment while still can be (indirectly) interpreted as binary (space / obstacle) labels and ii) the fact that multiple large-scale and annotated datasets are available~\cite{cityscapes, coco, zhou2017scene}.

In contrast to microphone arrays, cameras suffer from limited field-of-view (except for omnidirectional cameras).
Thus, we need to process multiple images to ``understand'' the whole scene (\eg a room), which unfortunately introduces two major difficulties: i) predictions from independently processed images are often inconsistent \cite{sengupta2012iros, badrinarayanan2015segnet} and ii) such data has to be projected into a common representation (coordinate frame) shared with microphone array.
While the former can be suppressed by explicit data association~\cite{miksik2013icra, kundu2016cvpr}, the latter is typically addressed by projecting the data on a common representation such as semantic maps~\cite{sengupta2013icra, valentin2013cvpr, Hermans14ICRA, vineet2015icra}.

\section{Approach}
\label{sec:approach}

\subsection{Incorporation of static visual priors into DoA}
\label{sec:doa_with_prior}

In this letter, we consider the DoA methods that rely on optimising some objective w.r.t. a set of candidate solutions. Let us denote a \emph{discrete} set of potential angular directions $\varphi$ of a circular array as $\calG = \{\varphi_i \; | \; 0 \leq  \varphi_i \leq  2\pi, \; i \in \bbN \}$, the expected DoA is then assumed to be at $\varphi$ for which the cost function $f(\cdot)$ is maximised (or minimised). For example, for SRP-PHAT~\cite{dibiase2000}, one could search through all candidates in $\calG$, and select the one with the tallest peak, \ie
\begin{equation}
  \varphi^* = \argmax_{\varphi \in \calG} f(\varphi)
  \label{eq:doa_search}
\end{equation}
Similarly, we define static visual prior information as another set $\calP = \{\varphi_i, \; Z_i  \; | \; 0 \leq  \varphi_i \leq  2\pi, \; Z_i \in \{0,1\}, \; i \in \bbN\}$, in which $Z$ maps a given angular direction $\varphi_i$ into category of \emph{free} $(Z = 1)$ or \emph{obstructed} $(Z = 0)$ space. Then, one can alter the original search space $\calG$ with an additional information in $\calP$, \ie filter out directions that have been annotated as an obstructed space, and use $\calG^{'}$ in~\eqref{eq:doa_search} instead of $\calG$:
\begin{equation}
  \calG^{'} = \calG\{\cdot\} \cap \calP\{\cdot,Z = 1\}
  \label{eq:grid}
\end{equation}
Note that~\eqref{eq:grid} is independent of particular DoA model, thus the underlying characteristics behind back-end DoA estimator remain unchanged. 
%
Priors are estimated separately, hence can be easily swapped given the new $\calP$ is available (\eg when device was moved to a different location).
Also, visual prior linearly decreases computational cost for grid-based search algorithms, as one does not need to evaluate irrelevant directions when searching for the peaks. %
Finally, although we consider circular microphone arrays, our approach is applicable to arbitrary geometries assuming $\calG$ and $\calP$ were estimated correctly.

\vspace{-0.1cm}
\subsection{Building visual-based static spatial prior}

%
We draw inspiration from biological systems which use cognitive maps of the environment for decision making~\cite{brenden2016arxiv, dolan2013neuron, daw2005nature}. 
This is known in robotics as semantic maps and is typically built using \emph{semantic} Simultaneous Localisation and Mapping (SLAM)~\cite{davison2018spatialai}.
Such maps exist in various forms ranging from sparse symbols, through semi-dense and/or layered representations to fully-dense metric 3D maps.

We opt for \emph{layered panoramic} representation since it does not require specific sensors such as Kinect/stereo-cameras or data-driven depth estimation and provides reliable predictions even at large distances (tens of meters); hence it is considerably simpler and faster to build than dense metric 3D maps. 
At the same time, DoA typically does not estimate proximity of the sound source, therefore omitting \emph{depth} from spatial prior is not overly restrictive.
The fact we use only passive monocular camera is of paramount importance for practical applications as it is the most widely used imaging sensor already commonly available on majority of potential hardware platforms such as Amazon Echo Spot/Show2~\cite{echospot} or mobile phones.
Given a set of input images $\calI = \{I_1, I_2, \ldots, I_n\}$, our goal is to output a spatial prior map expressed as continues angular representation $\calP = \{\varphi, Z\}$ where $\varphi$ is an angle and $Z \in \{0,1\}$ denotes obstacles and free space (\cf \S\ref{sec:doa_with_prior}). 
We assume that all input images $\calI$ were collected using a sensor spinning around its vertical axis 
(\cf \S\ref{sec:experiments}) 
so they (approximately) share the same camera center and hence induce a homography~\cite{Hartley2004}.
In other words, we can estimate a homography matrix $\bfH_{i,j}$ representing a 1-to-1 mapping (warp) between any pair of sufficiently overlapping images $i$ and $j$. 
We use a standard approach of~\cite{brown2007panorama} which takes a set of input images $\calI$ and outputs a set of corresponding pairwise homography matrices $\calH = \{\bfH_{1,2}, \bfH_{2,3}, \ldots, \bfH_{n, 1}\}$, using sparse feature matching, robust RANSAC-based homography fitting and bundle adjustment.
Rotation matrices $\bfR_{i,j} \in SO(3)$ can be extracted from homography $\bfH_{i,j}$ using \eg~\cite{malis:inria-00174036}.
To avoid time-consuming building of image dataset (computer vision models typically require tens thousands of labelled images~\cite{cityscapes, coco, zhou2017scene}), we predict \emph{free space} and \emph{obstacles} indirectly, using semantic segmentation.
This allows to use existing large-scale datasets, such as ADE20K~\cite{zhou2017scene} (consisting of $20210$ training images labelled with per-pixel ground-truth).
To this end, we learn a nonlinear function $f_\theta: I \rightarrow S$ mapping image $I \in \bbR^{w \times h \times 3}$ to output $S \in \bbR^{w \times h \times L}$.
Here, each pixel of output $S$ represents an $L$-dimensional scores vector corresponding to $L$ semantic classes and $w$ and $h$ are image dimensions.
The multi-class predictor $f$ is implemented as a convolutional neural network (CNN) and parametrized \mbox{by $\theta$}.
Finally, the output scores are mapped into binary labels (free space / obstacles) and projected into spatial prior map $\calP$ using a corresponding homography matrix $\bfH_{i, j}$.

\begin{figure}[!h]
    \centering
    \subfloat[]{\includegraphics[width=0.27\linewidth]{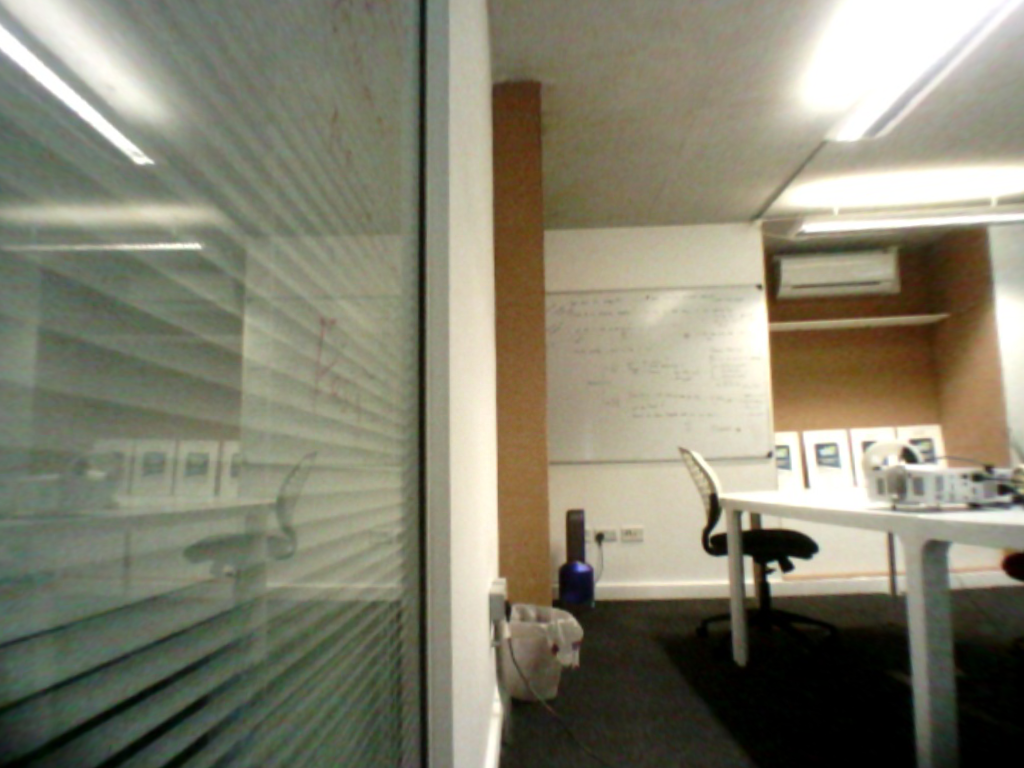}
                \includegraphics[width=0.27\linewidth]{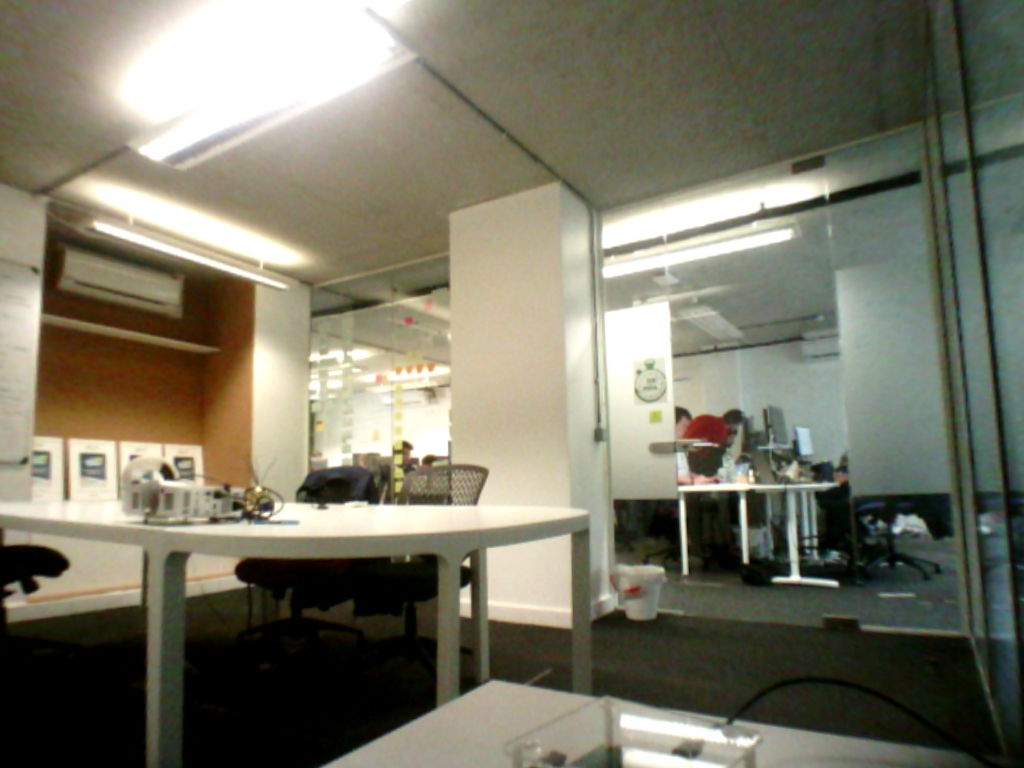}
                \includegraphics[width=0.27\linewidth]{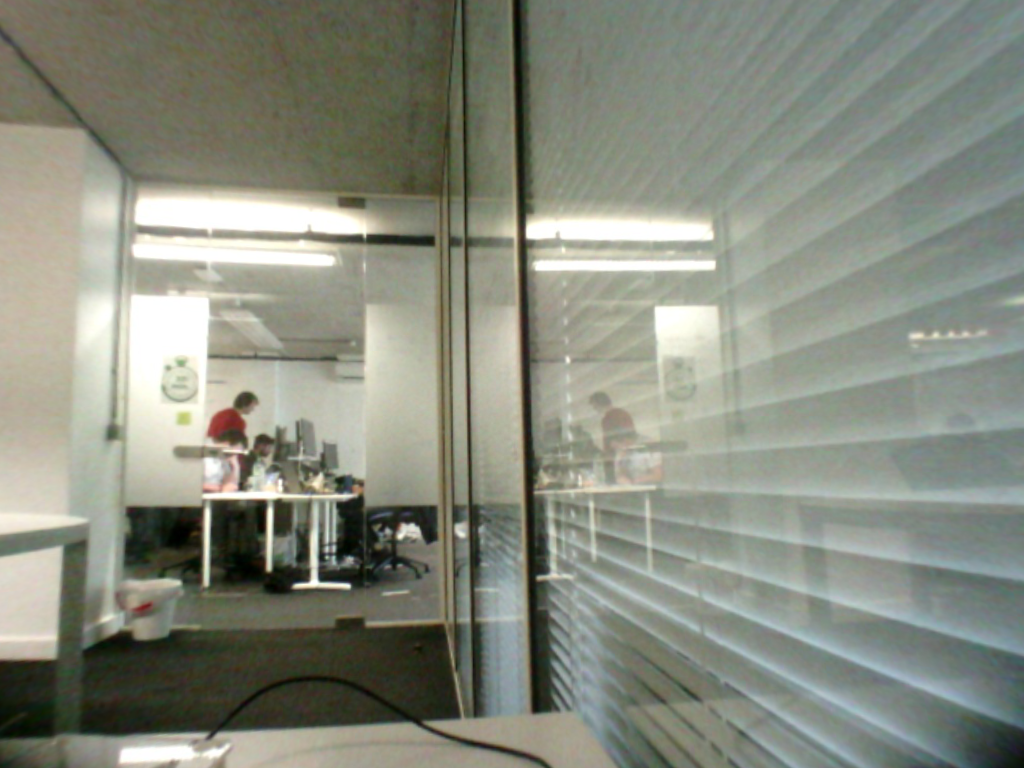}} \\ \vspace{-0.3cm}
    \subfloat[]{\includegraphics[width=0.38\linewidth, height=2cm]{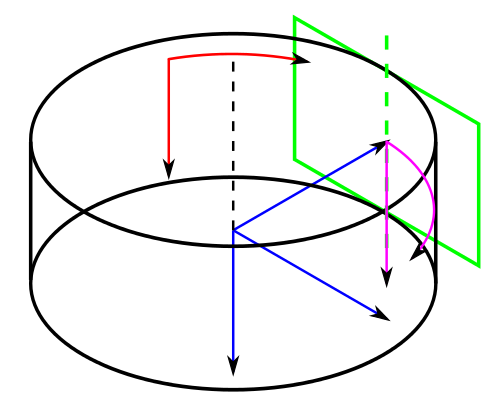}} \qquad
    \subfloat[]{\includegraphics[width=0.38\linewidth, height=2cm]{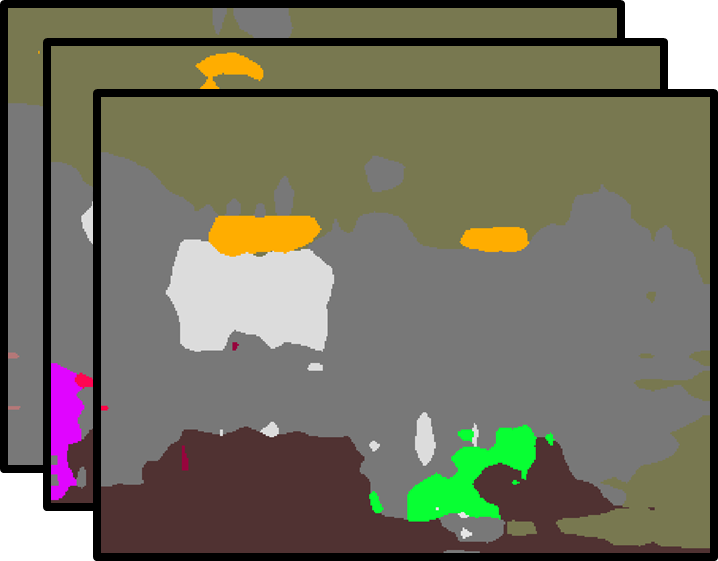}} \\ \vspace{-0.2cm}
    \subfloat[]{\includegraphics[width=0.85\linewidth, height=2cm]{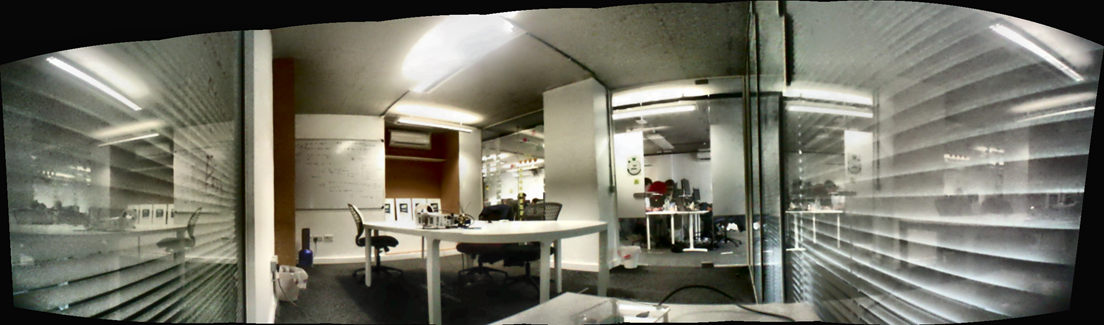}} \\ \vspace{-0.2cm}
    \subfloat[]{\includegraphics[width=0.85\linewidth, height=2cm]{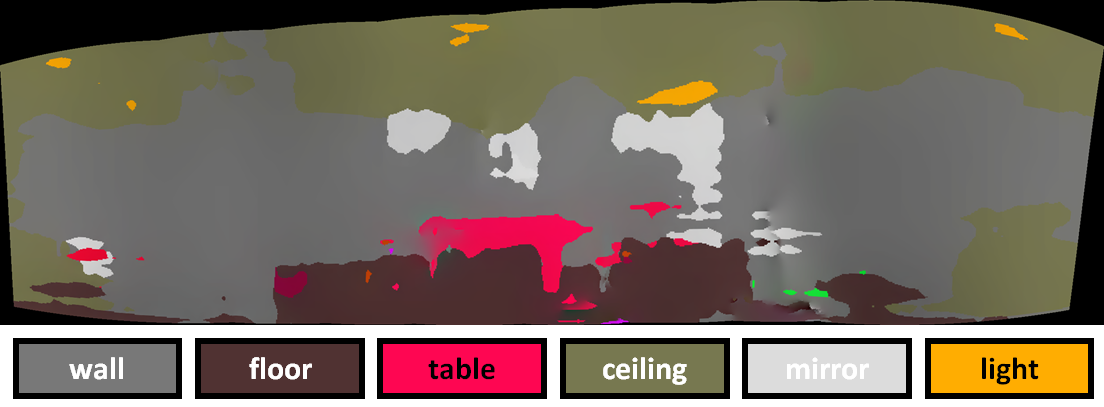}} \\ \vspace{-0.2cm}
    \subfloat[]{\includegraphics[width=0.85\linewidth, height=2cm]{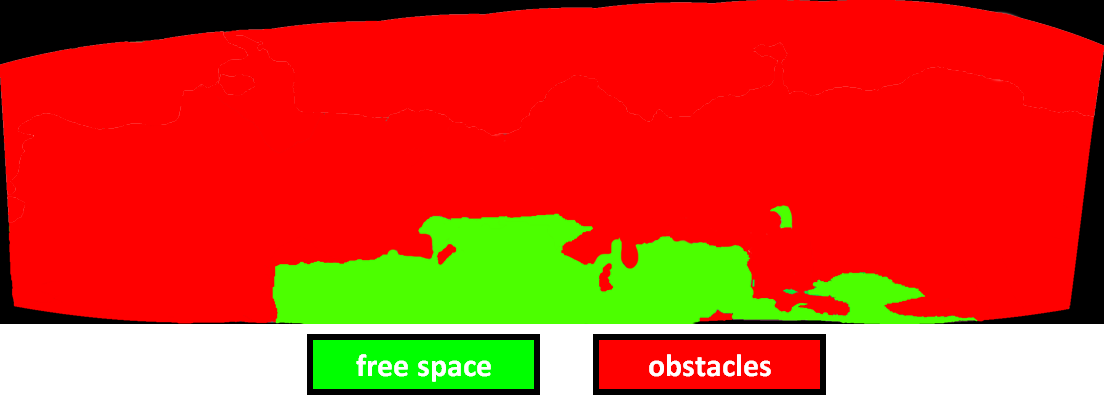}} \vspace{-0.15cm} 
\caption{Illustration of visual spatial prior construction. \mbox{(a) we} capture a set of images covering $360^{\circ}$ around the device. These images are in parallel used to (b) estimate homography matrices and (c) to predict semantic segmentation. This is then combined to built \emph{layered panoramic representation} (d) and (e), which is interpreted as \emph{free space} and \emph{obstacles} (f).}
\vspace{-0.35cm}
    \label{fig:building_spatial_prior}
\end{figure}

\textit{Implementation details.}
For panorama stitching, we adapted a public implementation of~\cite{brown2007panorama} from OpenCV to support semantic images. For semantic segmentation, we used the ADE20K Dataset~\cite{zhou2017scene} to train the \mbox{DilatedNet} model~\cite{YuKoltun2016} which drops \texttt{pool4} and \texttt{pool5} from fully convolutional VGG-16 network, and replaces the following convolutions with dilated (atrous) convolutions, and bilinear upsampling layer at the end.
%
%
Finally, to convert the labelling into the angular representation of \emph{free space} and \emph{obstacles} $\calP = \{\varphi, Z \}$, we check whether the fraction of pixels with semantic classes typically found in free areas (floor, ceiling, desk, chair, \ldots) within the current camera frustum is above per-class thresholds set using cross-validation on the \texttt{Dev} fold (\cf Appendix~\ref{sec:appC}).
%

\section{Experimental results} \label{sec:experiments}
\textbf{Experimental Protocol}. To the best of our knowledge, there is no publicly available dataset consisting of $360^{\circ}$ audio-visual data annotated with ground-truth directions of arrivals. Therefore, we collected around 2 hours of acoustic data with the corresponding visual snapshots representing office and home environments. \texttt{Dev} set comprises around 6 minutes of natural speech (collected in 2 office rooms, 15 sound source locations and 4 different microphone/camera placements). For the test set we collected two variants - \texttt{Test-Clean} comprising around 1h50min of re-recorded Librispeech data~\cite{Panayotov2015} as a primary sound source (2 rooms, 4 mic/camera placements, 5 sound-source and 2 noise-source locations). \texttt{Test-Noise} is a parallel variant (another 1h50min) collected in the identical conditions as \texttt{Test-Clean} but with present competing noise source (simulating a TV/radio) based on MUSAN data~\cite{musan2015}. We also include a fully synthetic control benchmark referred to as \texttt{Syn.}. 
Refer to the Appendix~\ref{sec:appA} for more details.

DoA predictions, unless stated otherwise, are computed  
on $256 \, \mathrm{ms}$ long analysis windows.
Signals are transformed to frequency domain with discrete Fourier transform, followed by an energy-based voice activity thresholding. 
DoA is calculated on frequency bins representing range from $500 \, \mathrm{Hz}$ to $8 \, \mathrm{kHz}$. DoA detection is carried with $180$ uniform bins, representing $2^{\circ}$ resolution ($|\calG|=180$). 
We used SRP-PHAT¬\cite{dibiase2000}, MUSIC~\cite{Schmidt1986} and TOPS~\cite{Yoon2006} as back-end DoA algorithms. SRP-PHAT was found to give the best results, thus in the remainder all analyses are based on this method (full results for all three techniques are reported in the Appendix~\ref{sec:appB}).
Experiments were carried out using an open source Pyromacoustics toolkit~\cite{scheibler2018pyroomacoustics}.

%
\textbf{Results}. Tab.~\ref{tab:results256} shows the main results for the three data folds, \texttt{Syn.}, \texttt{Dev} and \texttt{Test}. We report both average errors, as well as average bin accuracies defined as a percentage of predictions falling into a bin of assumed width on either side of the ground truth DoA ($\pm5$ degrees unless stated otherwise). 
DoAs with automatically extracted visual priors reduced average errors by an absolute $15.5^\circ$ and increased bin accuracy by 16.2\% relative (on \texttt{Dev} and \texttt{Test} sets, note CV priors are not available for \texttt{Syn.}). This effect was relatively stronger for \texttt{Test-Noise} condition, where automatically derived priors offered 25.9\% rel. bin accuracy increase.
Similar trends are observed with ground truth (expert) priors, where average degree errors were roughly halved in each of tested condition. Likewise, bin accuracies increased on average by 25.9\% rel. and this effect was stronger in \texttt{Test-Noise} variant at 41.8\% better vs. avg. 17.9\% for \texttt{Dev} and \texttt{Test-Clean}. Note that those numbers concern difficult cases such as a device next to the wall, or in the corner. In case where device is further away from the walls, baseline errors are lower at 10\% (\ie as for M3 position in \figref{fig:eagle_room} in the Appendix~\ref{sec:appA}).

\begin{table}[t] 
 \caption{Average error rates and $\pm$5 deg bin accuracies (in square brackets) for the synthetic, dev and test sets obtained with SRP-PHAT algorithm with and without spatial priors.}
\centering \small
 \begin{tabular}{|c|c||c||c|c|}
 \hline
            & \multicolumn{4}{c|}{ Avg. Error (deg) [$\pm$5 deg bin acc (\%)] } \\ \cline{2-5}  
            &   \texttt{Syn.}             &   \texttt{Dev}            &   \multicolumn{2}{c|}{\texttt{Test}} \\ \cline{4-5}
\bf{Prior}   & &   & \texttt{Clean} & \texttt{Noise} \\
 \hline\hline 
 None       & 3.2 [98.5] & 25.1 [52.6]   & 46.2 [52.9] & 73.7 [35.1]\\ \hline
 Visual   & N/A   & 14.5 [58.5]   &  31.1 [59.2]  &  52.7 [44.1] \\ \hline \hline
 Expert   & 1.6 [99.5] & 11.6 [62.5]   & 25.3 [61.9] & 39.3 [49.8]\\ \hline
 \end{tabular}
\label{tab:results256}
 \vspace{-0.3cm}
\end{table}

\begin{figure}[t]
    \centering
    \includegraphics[width=0.8\linewidth]{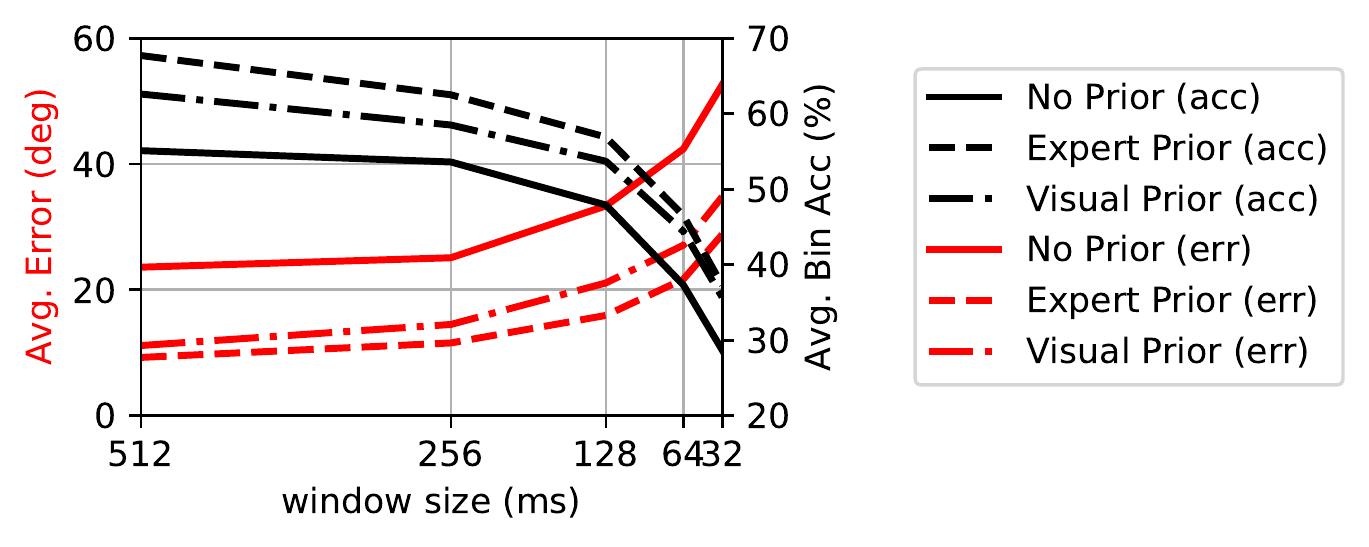}  \vspace{-0.4cm} \\
    \includegraphics[width=0.485\linewidth]{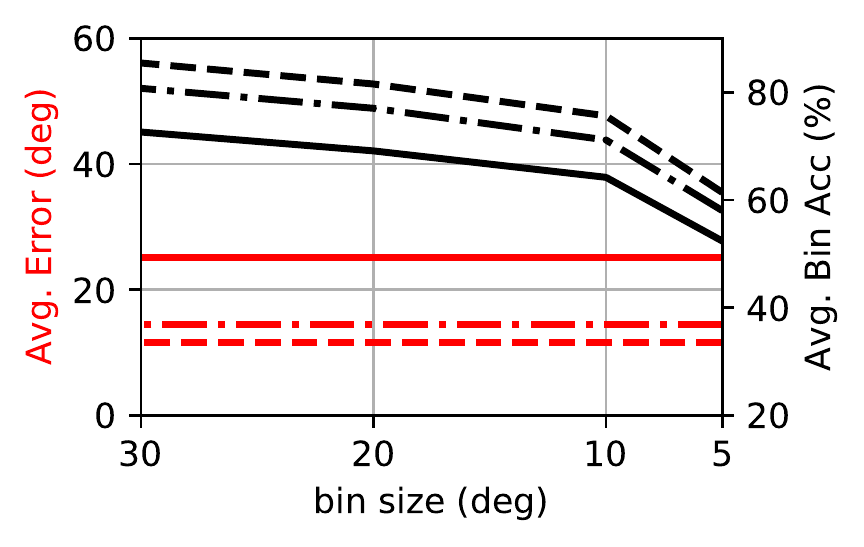} 
    \includegraphics[width=0.475\linewidth]{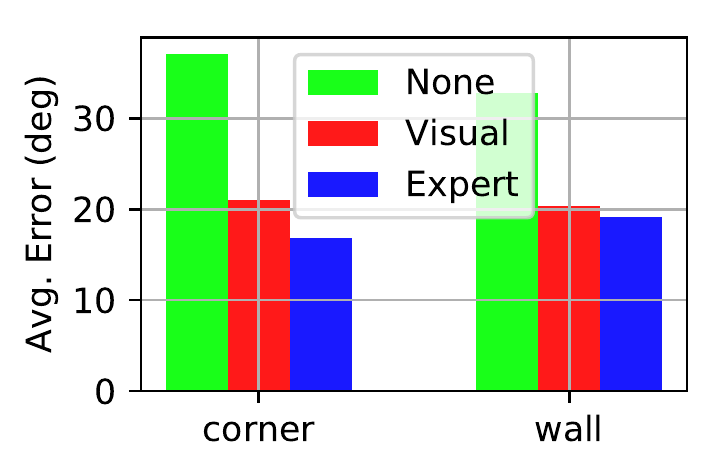}
    \vspace{-0.2cm}
    \caption{Effects of priors on avg. errors (red) and bin acc. (black) on \texttt{Dev} data as a function of (top) window sizes (left) bin sizes and (right) prior widths (\ie corner, wall). 
    }
    \label{fig:dev}
    \vspace{-0.2cm}
\end{figure}

\figref{fig:dev} offers more insight into other operating points on \texttt{Dev}. In particular, \figref{fig:dev} (top) shows average errors and bin accuracies for analysis windows ranging from $32 - 512 \, \mathrm{ms}$.
As expected, longer windows (more snapshots) offer lower errors and higher bin accuracy, though at the cost of latency (could be an issue for some applications like mapping acoustic events to spatial locations).
\figref{fig:dev} (left) shows how bin accuracies varies for different bin widths - this is of interest as even coarse prediction (\ie $\pm30$ deg bin) is still acceptable, as it may put the sound source within camera's field-of-view, which can be then used to fine-tune DoA~\cite{strobel2001joint, gehrig2005kalman, Debru2017}. In either scenario, trends are consistent and both expert and visual priors significantly outperform the baseline. \figref{fig:dev} (right) shows how errors varies for different prior widths (\ie corner, wall). Interestingly, tighter priors offer larger relative improvements and this effect increases for shorter analysis windows (\cf results in the Appendix~\ref{sec:appB}).
Finally, \figref{fig:doa_example} illustrates an example sequence of DoA prediction under all three settings.

\begin{figure}[t]
    \centering
    \includegraphics[width=1.0\columnwidth]{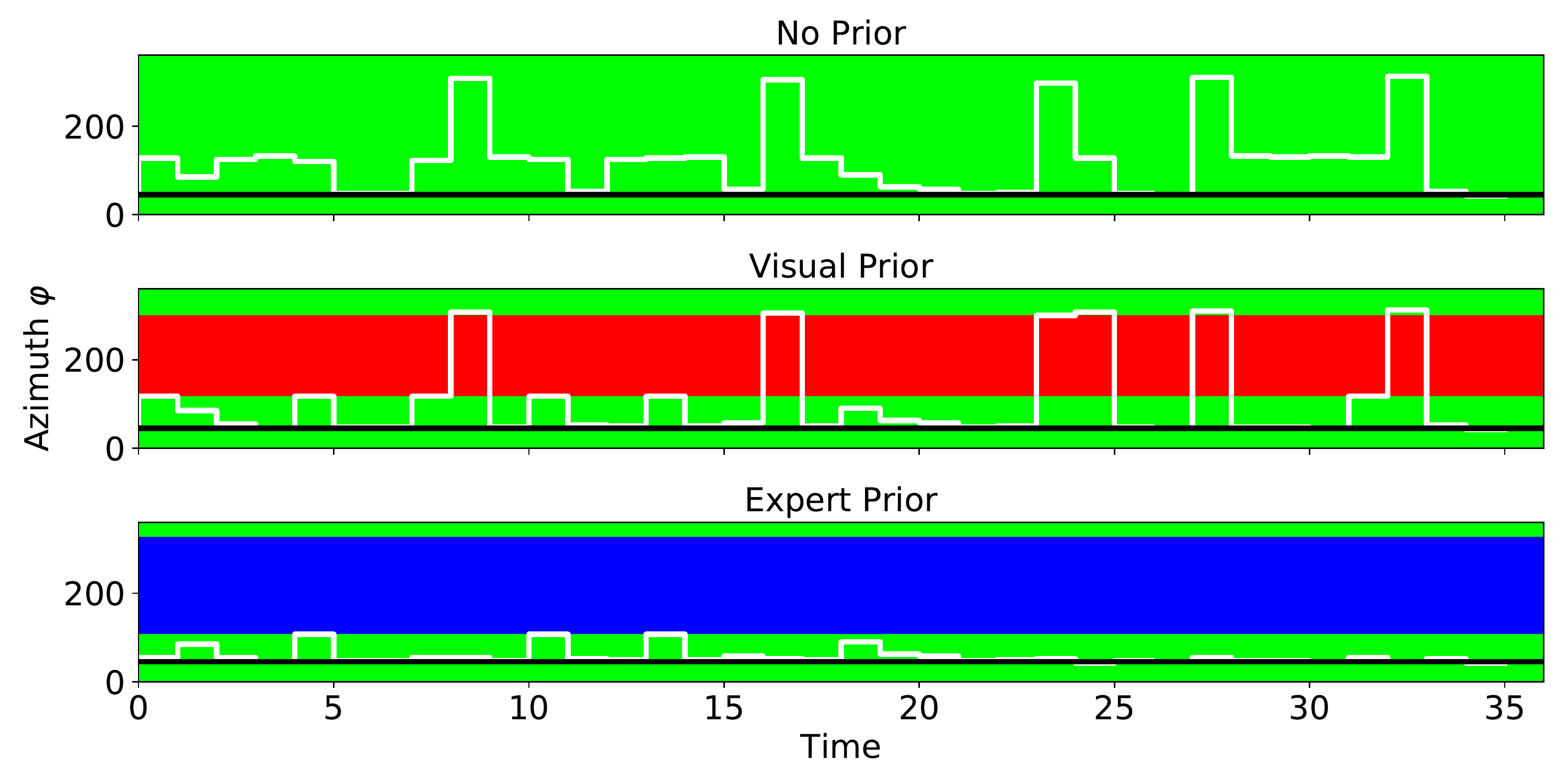}
    \vspace{-0.55cm}
    \caption {Visual priors for free space (green) and obstacles (red, blue) define allowed search regions for DoA algorithms.
    Solid black line is the ground-truth (at $45^{\circ}$), white lines denote predicted DoAs in analysis windows.
    (Top) Baseline DoA predictions without spatial priors. 
    (Middle) Estimated DoAs with automatically derived spatial prior using computer vision (obstacles between 120$^{\circ}$ and 300$^{\circ}$). 
    \mbox{(Bottom) The} expert derived ground-truth prior (obstacles between 110$^{\circ}$ and 320$^{\circ}$).
    %
    %
    This scene corresponds to the one depicted in the Fig.~\ref{fig:building_spatial_prior} (d) in which the device ``observes'' the room from the corner.}
    \label{fig:doa_example}
    \vspace{-0.4cm}
\end{figure}

\section{Discussion and Conclusions}


While our approach has demonstrated promising results,
there are many possible extensions. 
For instance, one could improve static priors by exploiting richer semantic information, \ie detect acoustically active noise sources such as TV or speakers and further constrain \emph{free} space to regions that are more likely to be occupied by people vs. devices (for talker localisation). 
Similarly, such priors could be used to discover and set nulls for known sources of acoustic interference in the generic sidelobe canceller class of beamformers~\cite{van2004optimum}.
Our layered panoramic representation is only an example of spatial prior map construction, however, it offers a number of interesting features as it is i) computationally and memory efficient, ii) continuous (\ie no angular quantization, temporal consistency) and iii) scalable (arbitrary number of classes, depth, materials, surface normals, \etc).
%
%
Experiments proved this is an efficient approach (\cf \S\ref{sec:experiments}), however, a variety of alternative approaches exists; for instance, one could learn a CNN to predict free space directly, instead of semantic segmentation (given appropriately annotated data).
Another option might be using dense metric 3D representation, if more advanced sensors are available.

We have proposed the first multi-modal DoA, which uses static visual spatial prior to reduce potential false detections.
We have validated our approach on a newly collected real-world dataset, and showed that our approach consistently improves over a wide range of DoA baselines using the ground-truth prior as obtained by an expert.
Finally, we have demonstrated a real-world performance of our approach using a simple method for deriving spatial prior automatically.

{
\small
\bibliographystyle{IEEEbib}
\bibliography{main}
}

\clearpage

\renewcommand{\thefigure}{A\arabic{figure}}
\renewcommand{\thetable}{A\Roman{table}}
\setcounter{figure}{0}
\setcounter{table}{0}

\begin{appendices}

\section {$\;$} 
\vspace{-0.5cm}
\subsection {Data Collection} \label{sec:appA}
We have built an experimental platform consisting of a fixed base and a moving head equipped with an RGB wide-angle camera, two $4$-microphone arrays with circular geometries\footnote{The 1st array (used to collect \texttt{Dev}) had $28.3$mm radius, the 2nd array (used to collect \texttt{Test}) was out-of-the-shelf XMOS VocalFusion XVF3100.}, and a DC motor allowing for continuous $360^{\circ}$ rotation around its vertical axis. 
The reference position of the moving head (with camera) was calibrated with respect to microphone array to allow mapping from visual to audio data.
%
Using this platform, we have collected realistic acoustic and visual data necessary to construct and test injection of static visual priors. 
%

\begin{figure}[h]
    \centering
    \includegraphics[width=0.7\linewidth]{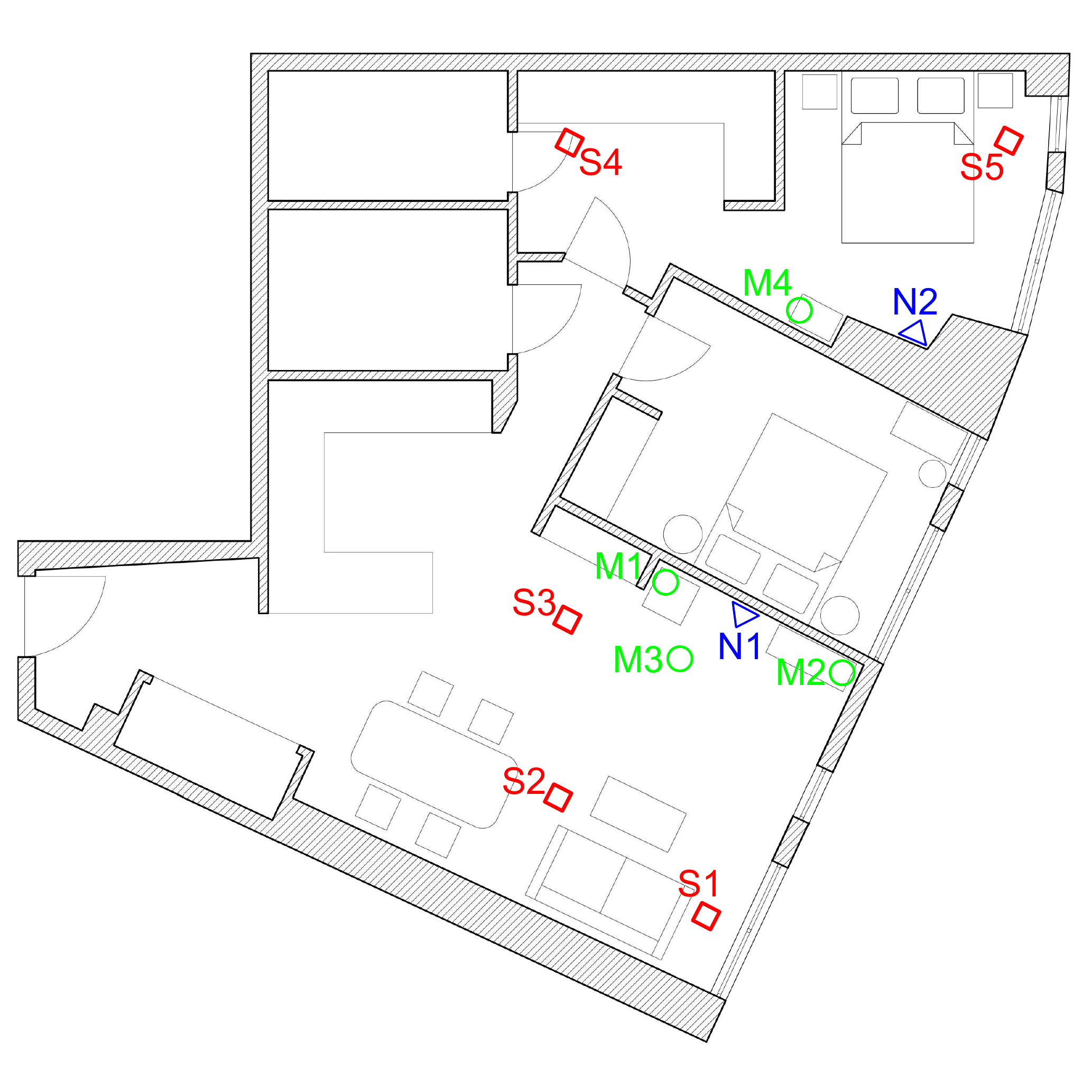}
    \vspace{-0.3cm}
    \caption{\texttt{Test} set collection environment with annotated positions of speakers (\textcolor{red}{S1-S5}), microphones (\textcolor{green}{M1-M4}) and noise sources (\textcolor{blue}{N1-N2}). Not to scale. }
    \label{fig:eagle_room}
    \vspace{-0.3cm}
\end{figure}

We split this data into development set \texttt{Dev} comprising natural speech with no competing sound sources, but in rooms characterised by high reverberation, with $T_{60}$ being approximately 600ms and 250ms., and test sets referred to as \texttt{Test-Clean} and \texttt{Test-Noise}.
Locations of sound and noise sources and mic/camera positions are shown in~\figref{fig:eagle_room}. Synthetic data \texttt{Syn.} is sampled from normal distribution. To simulate different impeding angles (6 in total) the source channel is shifted w.r.t. itself $M-1$ times using delay filter banks corresponding to the 1st mic array geometry. SNR augmented data is obtained by adding Gaussian noise at desired SNR levels separately to each of the shifted channels.

\vspace{-0.4cm}
\subsection {Mapping semantic segmentation to binary labels} \label{sec:appC}
To convert multi-class semantic segmentation to binary labels, we learnt per-class thresholds on the \texttt{Dev} fold, for the following classes \texttt{\{floor, desk, table, chair, tv\}} and assigned full images as \emph{free space} if the number of pixels with these classes is above threshold (recall the images are sampled with $10^\circ$ angular resolution). Of coarse, these are not the only available classes in the ADE20K Dataset, however, in practice worked well on the \texttt{Dev} fold and generalised to the test sets, as is demonstrated by presented results.
To convert recognised free space to angular representation, we simply read out images, which were assigned to \emph{free space} label as the images are captured relatively densely, however, one could in practice decompose the homography matrix, as described in \S\ref{sec:approach} and achieve much finer resolution if desired.

\vspace{-0.4cm}
\subsection {More results} \label{sec:appB}

Table~\ref{tab:results_alg} shows results for SRP-PHAT, MUSIC and TOPS on \texttt{Dev} set. All share the same automatically extracted visual and expert priors as well as identical pre-processing pipelines (the number of DFT points was optimised for each method).


Table~\ref{tab:results128} offers similar set of results to Table~\ref{tab:results256}, but for 128ms analysis window. The overall findings are consistent, but here priors injection offer larger relative gains across all sets (this trend increases as analysis window gets smaller). 

\begin{figure}[h!]
    \centering
    \subfloat[]{\includegraphics[width=0.5\linewidth]{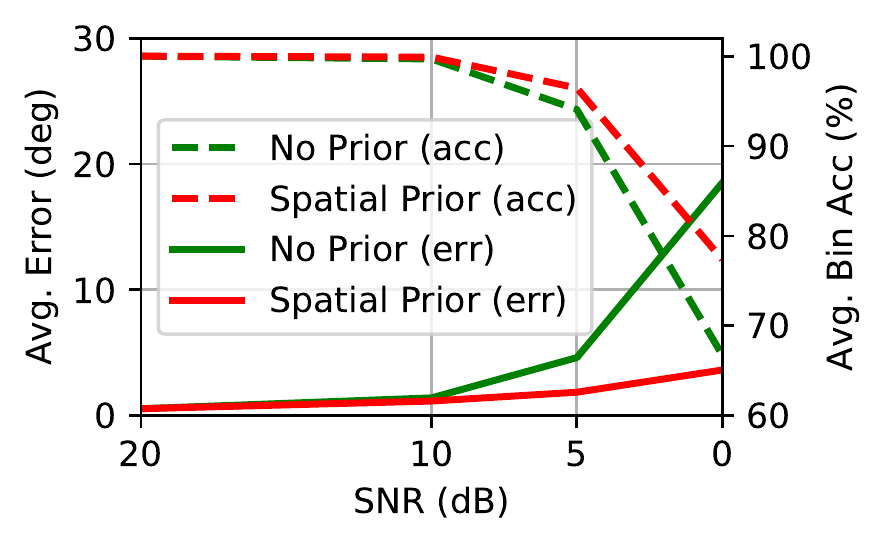}} 
    \subfloat[]{\includegraphics[width=0.5\linewidth]{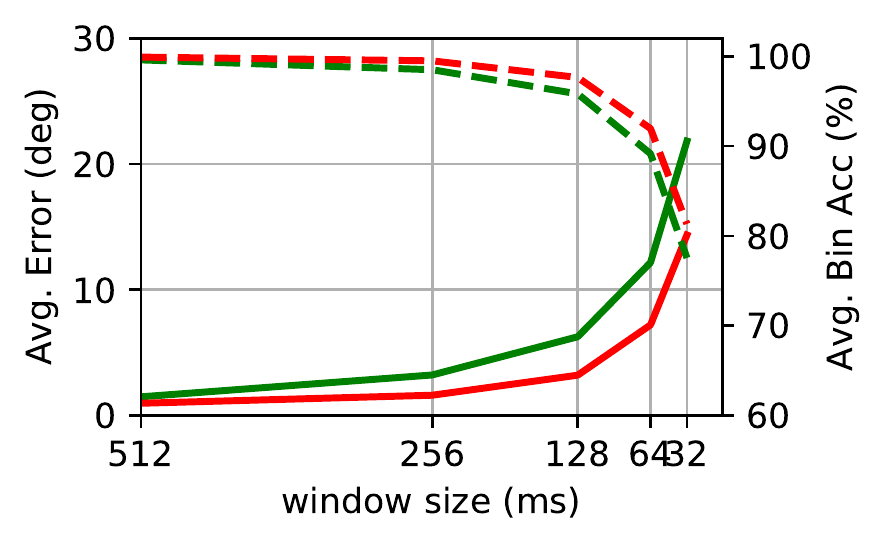}} \vspace{-0.4cm} \\
    \subfloat[]{\includegraphics[width=0.5\linewidth]{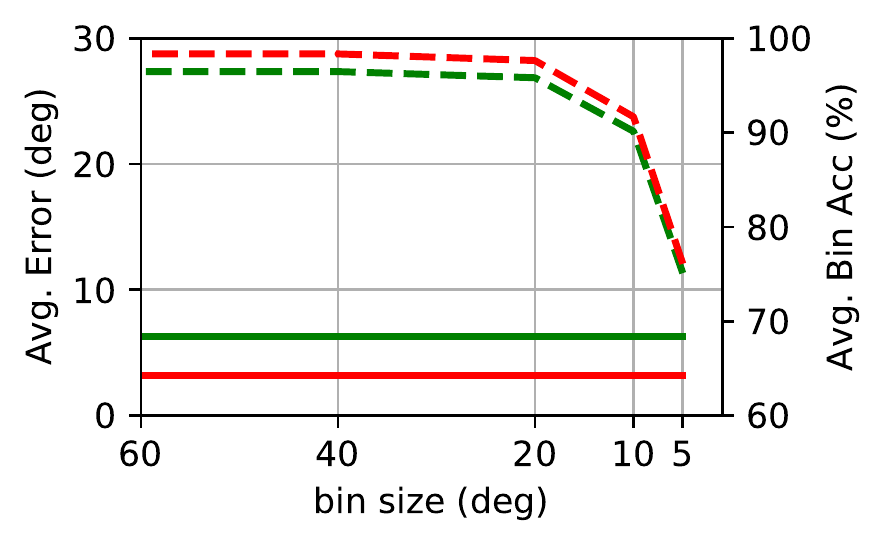}} 
    \subfloat[]{\includegraphics[width=0.5\linewidth]{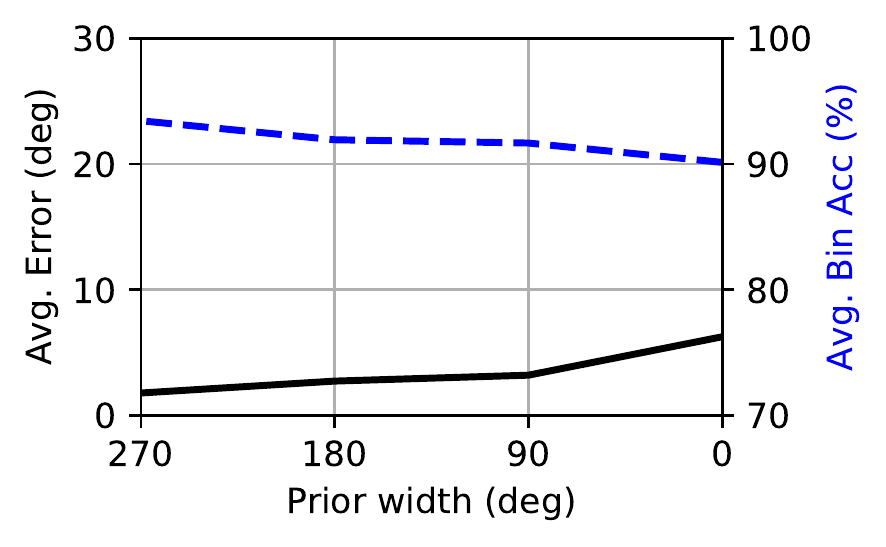}}
    \vspace{-0.2cm}
    \caption{Effects of prior on avg. errors (solid) and bin acc. (dashed) on \texttt{Syn.} data as a function of (a) SNRs (b) window sizes (c) bin sizes and (d) prior widths (i.e. corner, wall). Window $128$ms, average over scores of all SNR levels. 
    }
    \label{fig:synth}
    \vspace{-0.2cm}
\end{figure}

Finally, \figref{fig:synth} shows impact of expert priors on synthetic dataset. We test to what extent priors help under different SNR regimes~\figref{fig:synth} (a), analysis window and bins sizes (b) and (c) as well as prior widths (d), \ie $270^{\circ}$ approx. corresponds to a corner while $180^{\circ}$ to a wall case. Note, all errors on \texttt{Syn.} set are considerably lower when compared to realistic folds, thus gains are primarily visible in more challenging operating conditions (low SNR, short windows), however, overall findings are in-line with 
\texttt{Dev} and \texttt{Test} sets.

\vspace{-0.2cm}

\begin{table}[h] 
 \caption{Results on \texttt{Dev} set obtained with three DOA algorithms and considered priors for 128ms analysis windows.}
\centering \small
 \begin{tabular}{|c|c|c|c|}
 \hline
            & \multicolumn{3}{c|}{ Average Error (deg.) [$\pm5$ deg. bin acc. (\%)]} \\ \cline{2-4}  
\bf{Prior}   & \bf{SRP-PHAT}\cite{dibiase2000} &  \bf{MUSIC}\cite{Schmidt1986} & \bf{TOPS}\cite{Yoon2006} \\
 \hline\hline 
 None       &  33.4 [47.9]  & 56.9 [28.4]   & 38.5 [33.8] \\ \hline
 Visual       & 21.1 [54.2] & 33.7 [33.8]  & 18.2 [41.8] \\ \hline \hline
 Expert  & 15.9 [56.9] &  28.0 [34.1]  & 13.8 [42.3] \\ \hline
 \end{tabular}
\label{tab:results_alg}
 \vspace{-0.3cm}
\end{table}

\begin{table}[h] 
 \caption{Results for 128ms long analysis window and SRP-PHAT algorithm with and without spatial priors.}
\centering \small
 \begin{tabular}{|c|c||c||c|c|}
 \hline
            & \multicolumn{4}{c|}{ Avg. Err (deg) [$\pm$5 deg bin acc (\%)] } \\ \cline{2-5}  
            &   \texttt{Syn.}             &   \texttt{Dev}            &   \multicolumn{2}{c|}{\texttt{Test}} \\ \cline{4-5}
\bf{Prior}   & &   & \texttt{Clean} & \texttt{Noise} \\
 \hline\hline 
  None       & 6.3 [90.1] & 33.4 [47.9]   & 55.3 [44.0] & 77.8 [30.7]\\ \hline
 Visual   & N/A   & 21.1 [54.2]   & 36.4 [52.1]  &  54.7 [40.6] \\ \hline \hline
 Expert   & 1.8 [93.4] & 15.9 [56.9]   & 28.4 [55.8] & 41.4 [45.9]\\ \hline
 \end{tabular}
\label{tab:results128}
 \vspace{-0.4cm}
\end{table}

\end{appendices}

\end{document}